\documentstyle[aps,twocolumn,epsf]{revtex}

\def\be{\begin{equation}}
\def\ee{\end{equation}}

\begin{document}

\title{Trapping 
Reactions with Randomly Moving Traps: 
Exact Asymptotic Results
for Compact Exploration.
}

\author{G.Oshanin$^1$, O.B\'enichou$^2$, M.Coppey$^1$, and M.Moreau$^1$}

\address{$^1$ Laboratoire de Physique Th{\'e}orique des Liquides (CNRS - UMR 7600), 
Universit{\'e} Pierre et Marie Curie,
4 place Jussieu, 75252 Paris Cedex 05, France
}

\address{$^2$ Laboratoire de Physique de la Mati{\`e}re Condens{\'e}e,
Coll{\`e}ge de France, 11 place M.Berthelot, 75231 Paris Cedex 05, France
}

\address{\rm (Received: June 17, 2002)}
\address{\mbox{ }}
\address{\parbox{14cm}{\rm \mbox{ }\mbox{ }
In a recent Letter 
Bray and Blythe have shown 
that the 
survival probability $P_A(t)$ 
of an $A$ particle diffusing with a diffusion coefficient $D_A$ 
in a 1D system 
with diffusive traps $B$ 
is independent
of $D_A$
in the asymptotic limit $t \to \infty$ 
and 
coincides with the survival probability of 
an immobile target in the presence of diffusive traps.
Here we 
show that this remarkable behavior
has a more general range of validity and holds for systems 
of an arbitrary dimension $d$, integer or fractal,  provided that 
the traps are "compactly exploring" the space, 
i.e. 
the "fractal" dimension $d_w$ of traps' trajectories is greater than $d$. 
For the marginal case when $d_w = d$, as exemplified here by 
conventional diffusion in 2D systems, the 
decay form is determined up to a 
numerical factor in the characteristic decay time.
}}

\address{\mbox{ }}
\address{\parbox{14cm}{\rm PACS No:  05.40.-a, 02.50Ey, 82.20.-w}}
\maketitle

\makeatletter
\global\@specialpagefalse

\makeatother


Trapping $A + B \to B$ 
and recombination $A + B \to 0$ reactions (TR and RR)
involving randomly moving
$A$ and $B$ particles which 
react "when they meet" at a certain distance $b$ 
are ubiquitous in nature. A few 
stray examples
include quenching of 
delocalized 
excitations, 
coagulation, recombination of radicals, 
charge carriers or defects, 
or biological processes 
related to population survival
 \cite{rice}. 

In recent years there has been much interest 
in the long-time behavior
of these processes, following a remarkable  
discovery \cite{ov,burl,leb,bal,don,pastur,gp,kh,3} 
of many-particle effects, 
which induce
essential departures from the
conventionally expected behavior 
\cite{rice}.

A pronounced deviation from the 
text-book predictions was found
for the 
diffusion-controlled
RR
in case when initially 
the particles of the $A$ and $B$ species
are all distributed at random  
with strictly 
equal mean densities $n_0$. 
It has been first shown
 \cite{ov} and subsequently proven \cite{burl,leb} 
that as $t \to \infty$
the mean density $n(t)$ follows 
$n(t) \sim \alpha_d n^{1/2}_0 (D t)^{-d/4}$, 
where $d$ is the space dimensionality, 
$\alpha_d$ is a constant 
and $D = D_A + D_B$ is
the sum of particles' diffusion coefficients.  
This law 
contradicts the decay law obtained within the Smoluchowski approach (SA):
$n(t) \sim 1/\phi_b^{(d)}(t)$ \cite{rice}, where, as $t \to \infty$,
\begin{equation}
\label{Smol}
\phi_b^{(d)}(t) = \int_0^t d\tau K_S(\tau) \sim \displaystyle \left\{\begin{array}{lll}
\displaystyle 4 \sqrt{D t/\pi},  \;\;\;   \mbox{d = 1}, \nonumber\\
\displaystyle \frac{4 \pi D t}{\ln(4 D t/b^2)},  \;\;\;   \mbox{d = 2}, \nonumber\\
\displaystyle 4 \pi D b t,    \;\;\;    \mbox{d = 3},
\end{array}
\right.
\end{equation}
$K_S(\tau)$ being the $d$-dimensional Smoluchowski-type "constant",
defined as the flux of diffusive particles through 
the surface of an 
immobile sphere of radius $b$. 

For the TR  two 
situations were most thoroughly studied:
the case 
when $A$s diffuse while $B$s are static, 
and the situation in which the
$A$s are immobile while $B$s diffuse - 
the so-called target annihilation problem (TAP).
In the case
of static,  randomly 
placed (with mean density $\rho$)
traps the $A$ 
particle survival probability $P_A(t)$ shows
a non-trivial, 
fluctuation-induced behavior 
\cite{burl,leb,bal,don,pastur,gp,kh,3}
\begin{equation}
\label{traps}
\ln P_A(t) \sim - \rho^{2/(d + 2)} (D_A t)^{d/(d+2)}, \;\;\; t \to \infty,
\end{equation}
which is intimately related 
to many 
fundamenal problems of statistical physics 
\cite{burl,leb,bal,don,pastur,gp,kh,3,sosiska}.

Survival probability 
$P_{target}(t)$ of an immobile target $A$ of radius $b$ 
in presence of 
point-like 
diffusive traps $B$  (TAP) can be calculated exactly for any $d$
(see Refs.\cite{tach} and \cite{burl,blu,szabo}):
\begin{equation}
\label{k}
P_{target}(t) = \exp\Big( - \rho \phi_b^{(d)}(t)\Big),
\end{equation}
where $\phi_b^{(d)}(t)$ obeys  Eq.(\ref{Smol}) with $D_A = 0$. Decay forms in
systems with hard-core interactions between 
$B$s \cite{core} 
or with fluctuating chemical 
activity \cite{fluct} 
have been also derived. 

On contrary, the physically most important 
case of TR when both 
$A$s and $B$s diffuse
was not solved exactly. 
It has been proven \cite{leb} 
that here $P_A(t)$
obeys
\begin{equation}
\label{general}
\ln P_A(t) = - \lambda_d(D_A,D_B) \times \displaystyle \left\{\begin{array}{lll}
\displaystyle t^{1/2},  \;\;\;   \mbox{d = 1}, \nonumber\\
\displaystyle \frac{t}{\ln(t)},  \;\;\;   \mbox{d = 2}, \nonumber\\
\displaystyle t,    \;\;\;    \mbox{d = 3},
\end{array}
\right.
\end{equation}
which equation
defines its time-dependence exactly. On the other hand, 
the factor
$\lambda_d(D_A,D_B)$ remained as yet 
an unknown function of 
the particles' diffusivities and $d$. 
Since  the time-dependence of the 
function on the rhs of 
Eq.(\ref{general}) 
follows 
precisely the 
behavior of $\int^t d\tau K_S(\tau)$, 
one might  
expect that the 
SA provides 
quite 
an accurate description for this situation and 
following its spirit 
to set $D_A = 0$ supposing that 
traps diffuse with the diffusion 
coefficient $D = D_B + D_A$.  
As a matter of fact, it has been 
often tacitly assumed
that when 
both of species diffuse $P_A(t)$ obeys 
Eq.(\ref{k}) with 
$ \phi_b^{(d)}(t)$ defined by Eq.(\ref{Smol}) with $D = D_A + D_B$.  
On the other hand, 
it has been shown that $\lambda_d(D_A,D_B)$ is 
less than the corresponding prefactor 
in $K_S(t)$  \cite{burl} and 
that  it 
may be bounded by a non-analytic 
function of $D_A$ and $D_B$
\cite{bere}.
A perturbative approach for calculation
of $\lambda_d(D_A,D_B)$, as well as  
corrections to
the SA in 1D systems were presented \cite{szabo}. 
It has been also noticed 
that $\lambda_d(D_A,D_B)$ is not 
a function of 
$D = D_A + D_B$ only, since the diffusion-reaction equation 
are not separable \cite{szabo}.   
This lack of knowledge of the precise form  
of $\lambda_d(D_A,D_B)$, of course, constitutes
an annoying gap in the general 
understanding of the 
fluctuation phenomena in chemical kinetics. 

In a recent Letter Bray and Blythe (BB) \cite{bray}
made a considerable step 
towards the solution of this 
general case 
by showing that,
surprisingly, 
the 
survival probability of an $A$ particle diffusing 
in a 1D system 
with diffusive traps
is independent of $D_A$ 
in the asymptotic limit $t \to \infty$ 
 and 
coincides with the 
survival probability of 
an immobile target in 
the presence of 
diffusive traps, Eq.(\ref{k}). The convergence to 
this asymptotic
result might be, however, rather slow  as
shows 
the comparison \cite{bray} with extensive numerical simulations \cite{gras}. 

One may, however, 
pose quite a legitimate question whether 
such a remarkable result is 
constrained to  
physically not very realistic 
1D systems
 with conventional diffusion 
or if it is just a particular case 
of a more general behavior 
which persists also for higher $d$?

In this Letter we show that indeed 
this  
remarkable 
result 
holds  
for a more 
general case. Namely, we 
show that for systems 
of an arbitrary dimension $d$, integer or fractal, 
the large-$t$ behavior of the 
survival probability of 
a randomly moving $A$ 
particle in the 
presence of randomly moving traps
is given exactly by 
the solution of the exactly solvable 
TAP, provided that 
the traps are "compactly exploring" the space; in other words, 
the "fractal" dimension $d_w^{(B)}$ of 
traps' trajectories is greater than $d$ \cite{pgg}. 
For 
$\it lattice$ random walks, this corresponds to situations
in which random walks are 
recurrent \cite{hughes}. 
Random motion with $d_w^{(B)} > 2$ 
is widespread in nature and is most  
often encountered 
in porous and disordered systems, amorphous and
polymer materials \cite{hughes}, for which systems 
it will take place in two 
and even three dimensions (see, e.g.,  
Refs.\cite{pgg,mi} and \cite{mi2}). 
Finally, we examine the
behavior in the marginal 
case when $d_w^{(B)} = d$, as 
exemplified here by 
conventional diffusion in 2D systems, and show that 
here
 the 
decay form can be determined 
up to a 
numerical factor in the characteristic decay time.

Consider a $d$-dimensional volume $V$ containing
a single mobile $A$ particle of radius $b$  
and $N$ point-like traps $B$. 
Let ${\bf X}_t$ be the vector 
denoting the $A$ 
particle position
at time moment $t$, while 
${\bf x}^{(j)}_t$, $j=1, \ldots, N$, 
be the corresponding 
vector denoting the 
position of the $j$-th trap.
Introducing two auxiliary indicator functions
\begin{equation}
\delta_b({\bf x}) = \left\{\begin{array}{ll}
1,  \;\;\;   \mbox{$|{\bf x}| \leq b$}, \nonumber\\
0,    \;\;\;    \mbox{otherwise};
\end{array}
\right.
I(y) = \left\{\begin{array}{ll}
1,  \;\;\;   \mbox{$y = 0$}, \nonumber\\
0,    \;\;\;    \mbox{otherwise},
\end{array}
\right.
\end{equation}
one writes $P_A(t)$ down formally  
as follows:
\begin{equation}
\label{av}
P_A(t) = 
E\Big\{ \prod_{j=1}^N  \Big< I\Big(\int^t_0 \delta_b({\bf X}_\tau - {\bf x}_\tau^{(j)})
 d\tau\Big)\Big>_{\{{\bf x}_t^{(j)}\}} 
\Big\}, 
\end{equation}
where the symbol $E\{\ldots\}$ denotes 
averaging with respect to the A particle trajectories, while 
the brackets with the subscript $\{{\bf x}_t^{(j)}\}$ 
stand for averaging with respect to the trajectories of the 
the $j$-th trap. Note that
Eq.(\ref{av}) applies for any type of motion 
provided that the point-like traps are ignorant of each other and
thus move independently.  
In the limit $N,V \to \infty$ ($N/V = \rho$) one has 
\begin{equation}
\label{av1}
P_A(t) = 
E\Big\{ \exp\Big( - \rho
\int d {\bf x}_0 \Big< I'({\bf X}_t, {\bf x}_t )\Big>_{\{{\bf x}_t\},{\bf x}_{t=0} =
{\bf x}_0}
\Big)
\Big\},
\end{equation}
where brackets 
denote now averaging
with respect to the trajectories of a single trap $B$ whose
starting point is at position ${\bf x}_0$, while 
$I'({\bf X}_t, {\bf x}_t )$ 
is the indicator function
\begin{equation}
\label{av2}
I'({\bf X}_t, {\bf x}_t ) = 1 - I\Big(\int^t_0 \delta_b({\bf X}_\tau -
 {\bf x}_\tau)
 d\tau\Big),
\end{equation}
which shows whether two given realizations of 
trajectories ${\bf X}_t$ 
and ${\bf x}_t$ have
"intersected" each other at least once within 
the time interval $[0,t]$. 
Note that averaging over trajectories
${\bf x}_t$ in the exponential is to be taken 
for fixed ${\bf X}_t$, and
after performing such an averaging we have to 
average an exponential of the result 
with respect to the trajectories 
${\bf X}_t$, 
which represents a fairly complex mathematical
problem. Such a complexity emphasises, 
of course, the 
significance of the BB result.

Now, BB 
have noticed \cite{bray}, 
although not proven 
rigorously, 
that the $A$ particle will on average 
survive longer if it stays still than if it
diffuses. They have also furnished some arguments in favor of this statement showing
that this is true for
 systems
with a $\it finite$ number of 
traps since here
the lowest value of the 
decay exponent corresponds 
to $D_A = 0$.
In other words, 
it means that $P_A(t)$ in Eq.(\ref{av1}) 
is bounded from above by
\begin{equation}
\label{upper}
P_A(t) \leq P_{target}(t) = \exp\Big( - \rho \phi_b^{(d)}(t)\Big)
\end{equation}  
where  $\phi_b^{(d)}(t)$ is given by
\begin{eqnarray}
\label{volume}
&&\phi_b^{(d)}(t) =  \int d {\bf x}_0  
\Big< \Big[1 - I\Big(\int^t_0 \delta_b({\bf x}_\tau)
 d\tau\Big)\Big]  
\Big>_{\{{\bf x}_t\},{\bf x}_{t=0} = {\bf x}_0}  \nonumber\\
&=&   \int d {\bf x}_0  \Big< \Big[1 
- I\Big(\int^t_0 \delta_b({\bf x}_\tau - {\bf x}_0) d\tau\Big)\Big]  
\Big>_{\{{\bf x}_t\},{\bf x}_{t=0} = 0} 
\end{eqnarray}
Note also that Eq.(\ref{upper}) should hold 
for any type of 
random motion, not necessarily 
only for conventional diffusion.

A lower bound on $P_A(t)$ can be constructed in the following way 
\cite{bray}:
One notices first that
all terms in the product in Eq.(\ref{av}) 
are positive definite and hence, if one performs 
averaging of Eq.(\ref{av}) not over all 
$\it possible$ realizations of
trajectories ${\bf X}_t$ 
and ${\bf x}_t^{(j)}$, but only over some restricted 
subset, 
one arrives at a lower bound on $P_A(t)$. 
Following Ref.\cite{bray}, we define this
 subset as follows: let us suppose 
that initially the $A$ particle has been 
located at the origin while 
all traps were uniformly spread 
in a $d$-dimensional system such that the 
trap nearest to the origin 
appeared at distance $l$ from it.  
Then, we perform averaging 
only 
over such trajectories of the $A$ particles 
which do not leave 
within the time interval $[0,t]$ 
the volume  
of radius $l$ centered around the origin, 
and such trajectories of traps $B$ (which all 
initially are uniformly distributed 
 outside of this volume) 
do not enter there within the time interval $[0,t]$. 
For such trajectories
\begin{equation}
 \prod_{j=1}^N  I\Big(\int^t_0 \delta_b({\bf X}_\tau - {\bf x}_\tau^{(j)})
 d\tau\Big) \equiv 1,
\end{equation} 
and hence, the following lower bound is valid
\begin{eqnarray}
\label{lower}
P_A(t) &\geq& \exp( - V_d \rho l^d) \times {\rm Prob}(max\{|{\bf X}_\tau|\} < l| \tau \in [0,t]) \nonumber\\
&\times& {\rm Prob}_j(min\{|{ \bf x}_\tau^{(j)}|\} > l| \tau \in [0,t]),
\end{eqnarray}
where $V_d$ denotes the volume of a $d$-dimensional sphere of a unit radius, 
while two other multipliers stand
for the probability that the $A$ particle does not leave a sphere of radius $l$ 
within the time interval $[0,t]$ and the probability that neither of traps, initially 
uniformly distributed with mean density $\rho$ 
outside this sphere, enters this sphere up to time $t$. 
Note that exactly the same lower bound
has been already proposed in Refs.\cite{red2} and \cite{bere}.

Now, let us suppose that the mean-square displacement of the $A$ particle obeys
$<{\bf X}_t^2> \sim (D_A t)^{2/d^{(A)}_\omega}$, 
while the MSD of traps follows $<{\bf x}_t^2> \sim (D_B t)^{2/d^{(B)}_\omega}$,
$d^{(A)}_\omega$ and $d^{(B)}_\omega$ being the 
"fractal" dimensions  of the $A$ particle and traps trajectories, 
respectively. For conventional diffusion of has
$d^{(A)}_\omega = d^{(B)}_\omega \equiv 2$ for 
any $d$. Under quite general conditions, for $D_A t \gg l^{d^{(A)}_\omega}$
the probability ${\rm Prob}(max\{|{\bf X}_\tau|\} < l| \tau \in [0,t])$ 
can be estimated as \cite{hughes}
\begin{equation}
\label{p}
{\rm Prob}(max\{|{\bf X}_\tau|\} < l| \tau \in [0,t]) \sim \exp( -
 \beta_d (D_A t)/l^{d^{(A)}_\omega}), 
\end{equation}
where $\beta_d$ is a constant 
dependent on the type of random motion and $d$.
On the other hand, one readily notices that
${\rm Prob}_j(min\{|{ \bf x}_\tau^{(j)}|\} > l| \tau \in [0,t])$  
is just the probability that 
 an immobile 
target of radius $l$ 
survives until time $t$ in the presence of
randomly moving traps, i.e.
\begin{equation}
\label{n}
{\rm Prob}_j(min\{|{ \bf x}_\tau^{(j)}|\} > l| \tau \in [0,t])
= \exp\Big( - \rho \phi_l^{(d)}(t)\Big),
\end{equation}
where $\phi_l^{(d)}(t)$ is defined by Eqs.(\ref{volume}) with $b$ replaced by $l$.

We turn next to the most delicate point of our analysis.  
We note first 
that the 
definition in the first line in Eq.(\ref{volume})
allows to express $\phi_l^{(d)}(t)$, 
in virtue 
of the Gauss theorem, as a time-integral
 of $K_S(t)$ (see Ref.\cite{burl}). 
On the other hand,  the definition 
in the second line in Eq.(\ref{volume}) shows
that $\phi_l^{(d)}(t)$ can be thought of as
 the mean volume swept by 
randomly moving fictitious 
particle of radius $l$ during time $t$, i.e. 
the mean volume of the 
so-called "Wiener sausage" (see, e.g. Ref.\cite{bere}). Its lattice counterpart
is known as the mean number of distinct sites visited \cite{hughes}.
General properties of such a volume 
for different types of random motion
have been first
analysed
in the pioneering papers by de Gennes \cite{pgg}, in which he studied 
RR involving polymerized particles 
and TR on the percolation cluster.
As well, behavior of $\phi_l^{(d)}(t)$
 have been 
discussed at length in Ref.\cite{mi} 
within the context 
of the polymer-free voids distribution  in polymer solutions. 
It has been shown that
depending on the relation between $d^{(B)}_\omega$ and $d$, two  
completely different types 
of behavior may be observed. The first type of behavior
occurs when $d^{(B)}_\omega < d$. In this case $\phi_l^{(d)}(t)$, called 
by de Gennes as the "exploration volume", 
is smaller than the volume ${\bf x}_t^d$ where
the particle is confined. 
This case is called the case of non-compact exploration and
here   $\phi_l^{(d)}(t) \sim \gamma_d t$, where 
the prefactor $\gamma_d$ is some function dependent on $d$ and
the type of random motion. In this case $\gamma_d$ is
 $proportional$ to some positive 
power of $l$! 
For lattice random walks 
this regime takes place when random walks are non-recurrent \cite{hughes}.
In the opposite case
when $d^{(B)}_\omega > d$ 
the behavior is completely different. Here, the 
exploration volume  $\phi_l^{(d)}(t)$ 
increases sublinearly with time, 
the trajectories are spatially more confined
 and most of space inside 
the volume  ${\bf x}_t^d$ is 
indeed visited. 
This case is called the case of 
"compact exploration" 
(recurrent random walks for the lattice counterparts) 
and here 
\begin{equation}
\label{as}
\phi_l^{(d)}(t) \sim \Big(|{\bf x}_t|\Big)^d \sim (D_B t)^{d/d_\omega^{(B)}}, 
\;\;\; t \to \infty
\end{equation}
What is most 
important in this case is that the prefactor in this asymptotic law 
is $\it independent$ of 
 $l$! Note also that such a behavior is compatible with the Alexander-Orbach result
 $\phi_l^{(d=d_f)}(t) \sim  t^{d_f/d_\omega^{(B)}}$ 
for anomalous 
random walk on fractal lattices of dimension $d_f$ \cite{alo}.

Now, we note that the function 
on the rhs of Eq.(\ref{lower})
is valid for any 
value of $l$ and consequently, the 
"best" lower bound would 
correspond to such 
$l$ which provides 
its maximal value.
Focussing next solely
 on the case of  random motion 
with compact exploration, we note that in this case  
the leading large-$t$ behavior of $\phi_l^{(d)}(t)$ is  
independent of $l$,
and hence, 
we have to maximize only the 
product of the first two 
terms. This yields
\begin{equation}
\label{eq}
P_A(t) \geq 
\exp\Big(- \alpha_d' \; \rho^{1-z} (D_A t)^{z} \Big)
\; \exp\Big( - \rho \phi_l^{(d)}(t)\Big)
\end{equation}
where $\alpha_d'$ is a constant, 
$ z = d/(d + d_\omega^{(A)})$ and 
the asymptotic behavior 
of $\phi_l^{(d)}(t)$ is defined in Eq.(\ref{as}).

For compact exploration we have that $d < d_\omega^{(B)}$.  On the other hand, 
on comparing the growth rate in the exponent in the 
first term on the rhs of Eq.(\ref{eq})
against the growth rate 
of $\phi_l^{(d)}(t)$ defined in Eq.(\ref{as}), we infer 
that the second multiplier 
determines the overall 
decay in case when $d$, $d_\omega^{(A)}$ and  $d_\omega^{(B)}$ 
obey: $d < d_\omega^{(B)} < d + 
d_\omega^{(A)}$, which reduces to simple 
condition of compact exploration for  $d_\omega^{(A)} = d_\omega^{(B)}$. Further on, 
for $d$, $d_\omega^{(A)}$ and  $d_\omega^{(B)}$ which obey 
the double side inequality, we have evidently that 
the leading terms in Eqs.(\ref{upper}) and (\ref{eq}) 
coincide, since $\phi_b^{(d)}(t)$ is asymptotically   
independent of $b$ and $\phi_l^{(d)}(t)$ does not depend on $l$. We infer 
thus that in this quite general case
exact asymptotic solution for 
trapping reactions $A + B \to B$ in which both species move randomly
is given by the solution of 
the corresponding immobile 
target annihilation problem, which represents 
a substantial generalization of the BB result \cite{bray}.   

Finally, we analyze
the behavior in the 
marginal case $d = d_\omega^{(B)} = d_\omega^{(A)} $ using as an example 
conventional diffusion in 2D systems. In this case, an asymptotical behavior of 
$\phi_l^{(d=2)}(t)$ is well-known \cite{ber}:
\begin{eqnarray}
\phi_l^{(2)}(t) &=& \frac{4 \pi D_B t}{\Big(\ln(4 D_B t/l^2) - 2 \gamma\Big)} \Big[1 
+ \frac{A_1}{\Big(\ln(4 D_B t/l^2) - 2 \gamma\Big)} +  \nonumber\\
&+& {\cal O}\Big(\Big(\ln(4 D_B t/l^2) - 2 \gamma\Big)^{-2}\Big)\Big],
\end{eqnarray}
where $A_1 \approx 0.423$ and $\gamma$ is the Euler constant. 
On the other hand, for standard diffusive motion
one has that ${\rm Prob}(max\{|{\bf X}_\tau|\} < l| \tau \in [0,t])$ 
obeys Eq.(\ref{p}) in which one sets 
$d_\omega^{(A)} = 2$ and $\beta_2$ is the square of the first 
zero of the Bessel function $J_0(x)$.
Now, since $\phi_l^{(2)}(t)$ is only weakly (logarithmically) 
dependent on $l$, one may assume that 
the value of $l$ which maximizes the lower bound for $D_A t \gg l^2$ is still determined 
by the derivative of the first two terms on the rhs 
in Eq.(\ref{lower}), i.e. $l \sim (\beta_2 D_A t/V_2 \rho)^{1/4}$. 
Such an estimate shows then
 that in 2D systems 
with conventional diffusion 
the $A$ particle survival probability is 
bounded by:
\begin{eqnarray}
1 &+& \frac{A_1}{\ln(4 D_B t/b^2)} + {\cal O}\Big(\ln^{-2}(t)\Big) \; \leq \nonumber\\
&\leq& \; \ln\Big(1/P_A(t)\Big)  \frac{ \ln(4 D_B t/b^2)}{4 \pi D_B t \rho} \; \leq \; 2 - \nonumber\\
&-& 2 \; \frac{\Big(\ln(4/\beta_2) + \ln(V_2 \rho b^2)  + \ln(D_B/D_A)\Big)}{\ln(4 D_B t/b^2)} + 
{\cal O}\Big(\ln^{-2}(t)\Big) \nonumber
\end{eqnarray}
Hence, in this marginal 
case 
the 
suitably extended upper and lower 
bounds 
determine the 
decay form up to a 
numerical factor in the characteristic decay time.

The authors acknowledge helpful 
discussions with Professors Bray and Blythe on
the matters of this paper.

\end{document}